\documentclass[conference]{IEEEtran}
\pdfoutput=1

\usepackage{amsmath}
\usepackage{algorithmic}
\usepackage[linesnumbered]{algorithm2e}
\usepackage{color}

\usepackage[hyphens]{url}
\usepackage{breakurl}

\usepackage[usenames,dvipsnames]{xcolor}

\usepackage{graphicx}
\usepackage{listings}
\usepackage{setspace}
\usepackage{caption}
\usepackage{comment}
\usepackage{floatrow}
\usepackage{subfigure}
\usepackage[most]{tcolorbox}

\usepackage{endnotes,microtype,xspace,graphicx,fancyvrb,multirow}
\usepackage{fancyhdr}
\usepackage[normalem]{ulem}
\usepackage[hyphens]{url}

\newcommand{\subparagraph}{}
\usepackage[compact]{titlesec}
\titleformat{\paragraph}[runin]{\vspace{-3pt}\normalfont\normalsize\bfseries\small\sffamily
  }{}
  {0pt}{}[.]
  
\titleformat{\subparagraph}[runin]{\normalfont\normalsize\itshape
 }{}
 {0pt}{}{}

\newcommand{\todo}[1]{\textcolor{red}{TODO: #1}}

\newcommand{\insertFigure}[2]{
    \begin{figure}[t]
\setlength{\abovecaptionskip}{0pt}
\setlength{\belowcaptionskip}{0pt}
        \centering
        \includegraphics[width=\linewidth]{figs/#1.pdf}
        \caption{\small #2}
        \label{fig:#1}
    \end{figure}
}

\newcommand{\insertWideFigure}[2]{

    \begin{figure*}[h]
\setlength{\abovecaptionskip}{0pt}
\setlength{\belowcaptionskip}{0pt}
        \centering
        \includegraphics[width=\textwidth]{figs/#1.pdf}
        \caption{\small #2}
        \label{fig:#1}
    \end{figure*}

}

\newcommand{\squishlist}{
 \begin{list}{$\bullet$}
  { \setlength{\itemsep}{0pt}
     \setlength{\parsep}{3pt}
     \setlength{\topsep}{3pt}
     \setlength{\partopsep}{0pt}
     \setlength{\leftmargin}{1.5em}
     \setlength{\labelwidth}{1em}
     \setlength{\labelsep}{0.5em} } }

\newcommand{\squishlisttwo}{
 \begin{list}{$\bullet$}
  { \setlength{\itemsep}{0pt}
     \setlength{\parsep}{0pt}
    \setlength{\topsep}{0pt}
    \setlength{\partopsep}{0pt}
    \setlength{\leftmargin}{2em}
    \setlength{\labelwidth}{1.5em}
    \setlength{\labelsep}{0.5em} } }

\newcommand{\squishend}{
  \end{list}  }

\newcommand{\Sec}[1]{Section~\ref{#1}}

\newcommand{\ignore}[1]{}
\newcommand{\tool}{\textsc{SCALE-Sim}\xspace}

\title{SCALE-Sim: Systolic CNN Accelerator Simulator}
\author{
  \IEEEauthorblockN{
        Ananda Samajdar$^g$ \hspace{10mm}
        Yuhao Zhu$^r$ \hspace{10mm}
        Paul Whatmough$^a$ \hspace{10mm}
        Matthew Mattina$^a$ \hspace{10mm}
        Tushar Krishna$^g$ \hspace{10mm}
  }
  \IEEEauthorblockA{
    $^g$\textit{Georgia Tech} \hspace{10mm}
    $^r$\textit{University of Rochester} \hspace{15mm}
    $^a$\textit{ARM Research, Boston}
    \\
    \hspace{-15mm}anandsamajdar@gatech.edu
  }
}
\begin{document}
\maketitle
\pagestyle{plain}

\begin{abstract}

%

Systolic Arrays are one of the most popular 
compute substrates 
within Deep Learning accelerators today, as they 
provide extremely high efficiency for 
running dense matrix multiplications.
However, the research community lacks tools to 
provide principled insights on both the design 
trade-offs and efficient mapping strategies 
for systolic-array based 
accelerators.

We introduce \textbf{S}ystolic \textbf{A}rray
 \textbf{S}imulator (\tool), which is a configurable 
systolic array based cycle accurate DNN accelerator simulator. 
\tool exposes various micro-architectural features as well as system integration
parameters to the designer to enable comprehensive design space exploration. 
This is the first 
systolic array simulator tuned for running DNNs to the best of our knowledge.

Using \tool, we conduct a suite of case studies and demonstrate the effect of bandwidth, dataflow and aspect ratio on the overall runtime and energy of Deep Learning kernels across vision, speech, text, and games. We believe that these insights will be highly beneficial 
to both computer architects and ML practitioners.

\end{abstract}

\section{Introduction}
\label{sec:intro}




Deep Neural Networks (DNNs) have become prevalent for tackling
many performance-critical and energy-constrained tasks over the last few years,
such as real-time object detection~\cite{yolo}, keyword spotting~\cite{kws}, and
robot motion planning~\cite{pathplanning}.
Amongst the various DNN topologies in use today, convolutional neural networks
(CNNs) are arguably the most common.
The end of performance scaling for CPUs, and the high power budgets of GPUs have led to a 
a deluge of custom DNN accelerators from 
academia and industry~\cite{diannao2014asplos,
eyeriss, scnn, dadiannao2014micro, tpu-isca}.

Among previously proposed DNN accelerator architectures, 
2D array architectures are a prominent
choice, as they allow for operand reuse in two dimensions~\cite{scnn,
eyeriss,tpu-isca}.
Of the various 2D array options, systolic array architectures are a natural
match to CNNs because the local shifting data movement naturally echos the inherent
dataflow of a native 2D convolution.
Systolic arrays can also efficiently handle matrix-matrix and matrix-vector operations 
that arise during DNN training and running LSTMs respectively.
The efficiency of systolic arrays comes from operand movement being purely 
local (neighbor to neighbor) which in turn provides high compute density (i.e., low area), 
low-energy, and 
simplified control.
Coupled with a carefully-designed memory hierarchy, systolic array architectures can
leverage the abundant data reuse in DNNs while keeping the processing elements
(PE) busy to provide high throughput. 
Due to these properties, systolic arrays have been widely deployed - 
including the Google TPU ASIC ~\cite{tpu-isca}, the Xilinx FPGA 
overlays xDNN~\cite{xdnn}, and academic proposals \cite{scnn}.

It is natural to assume that we will see many more accelerators being proposed
as new DNN architectures and use cases are identified in the near
future, which poses a large spectrum
of efficiency and performance demands on the underlying accelerator design.
This makes it imperative to quickly prototype architectural ideas and iterate
over different designs. However, the various architecture design parameters have
non-trivial interactions and thus lead to complex design decisions. In addition,
different DNN topologies also cast significant implications on the optimal
hardware architecture, requiring us to often co-design the accelerator architecture 
with the class of DNN kernels of interest (e.g., edge vs cloud/inference vs training).




In this work, we identify the key design parameters of systolic arrays, and 
reveal first-order insights about 
their interplay and respective contributions towards end-to-end performance 
and energy-efficiency.
Also, we focus our attention on the fact that 
accelerators need to work as a part of a larger
system; thus focusing on the integration aspects of the design with the rest of the system is necessary to 
understand limits on real-world performance and scalability.


\insertFigure{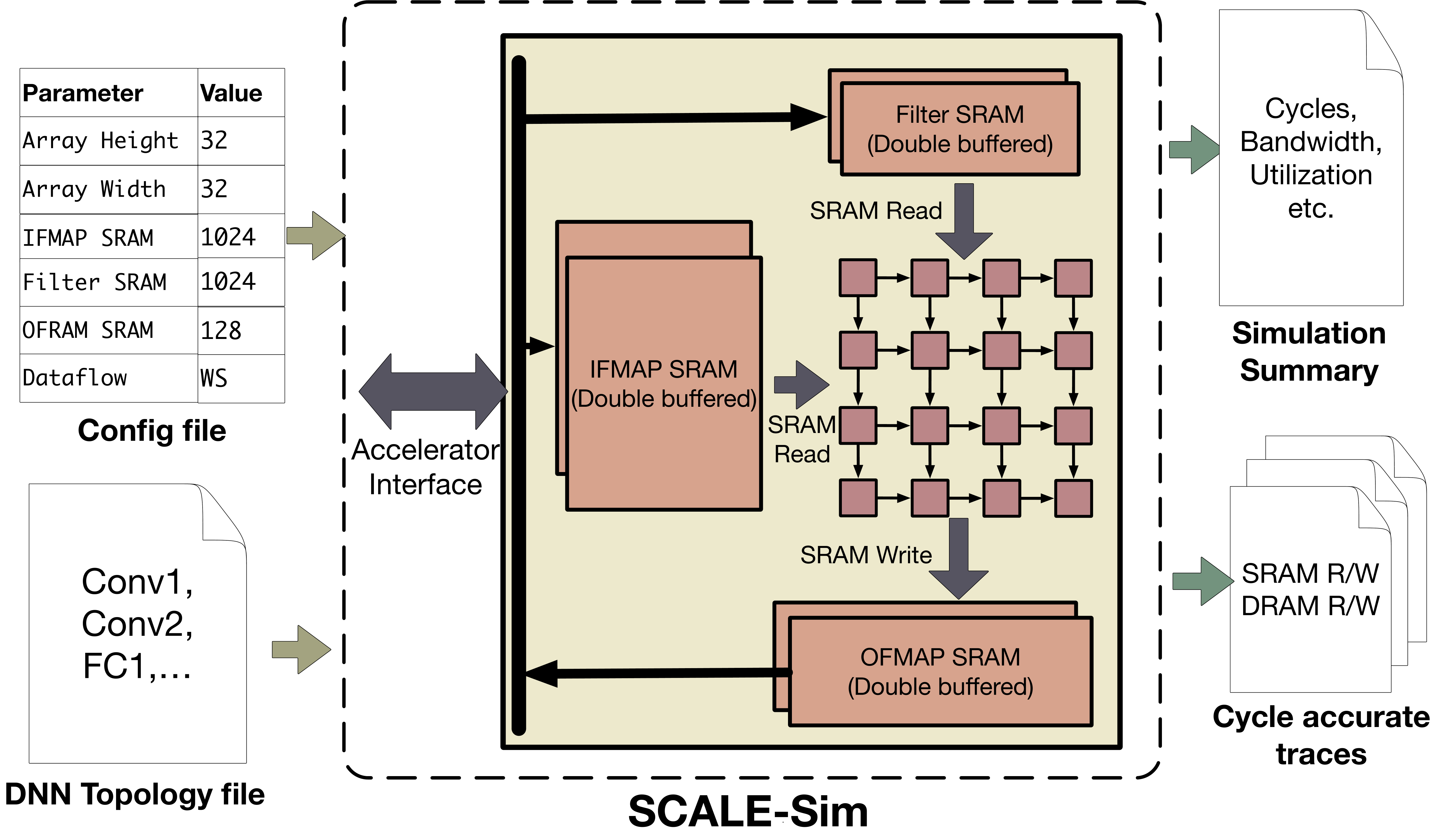}{Schematic depicting \tool, with inputs and outputs. The tool takes in architecture parameters as a config file, and the workload hyper-parameters as a csv file; and generates cycle accurate traffic traces and simulation summary csv files
}

With these goals in mind, we developed \tool (\textbf{S}ystolic \textbf{C}NN
\textbf{A}cce\textbf{le}rator \textbf{Sim}ulator),
cycle-accurate, systolic-array based CNN accelerator simulator. 
\tool exposes various micro-architectural features such as array size, array aspect ratio, scratchpad
memory size, dataflow mapping strategy, as well as system integration parameters
such as memory bandwidth.
Taking the microarchitectural parameters and the dimensions of each DNN layer as input, \tool 
reports the latency, array utilization, SRAM accesses, DRAM accesses, and 
DRAM bandwidth requirement.
\autoref{fig:scalesim} shows the schematic representation of the tool along with an example of 
its inputs and outputs.


Leveraging \tool, we perform a thorough analysis of the design space of the
systolic-array architecture, using the MLPerf benchmark \cite{mlperf}.
In the course of our study we identify several intricate trade-offs 
which were previously not thoroughly studied.
We report our findings in four categories. 
First, we study the affect of various dataflows over a fixed systolic array 
micro-architecture and report the trends we observe in performance and energy efficiency.
Second, we use \tool to identify the factors needed to be considered for properly sizing the on-chip scratchpad memories to extract most performance and energy efficiency from a design. 
Third, we study the affect of the array \textit{(aspect-ratio)} on performance for inference and report the trends. 
Finally, we perform a study to explore the trade-offs between two alternatives to increase performance, \textit{Scaling-Up} vs \textit{Scaling-out}.
Our experiments indicate that the micro-architectural parameters and workload hyper-parameters are closely intertwined and thus lead to interesting trends. For example, for non-square arrays certain networks perform better in taller array than wider given a particular dataflow. However, if the dataflow is changed, the same network starts performing well of wider arrays and runtime increases exponentially as the array is made taller. Also, while scaling compute, the performance improvements in Scaling-out vs Scaling-up solely depends on network hyperparameters, regardless of dataflow.


In summary, we make the following contributions:

\begin{itemize}
  \item We provide the first open-source, cycle-accurate DNN accelerator simulator based on the systolic-array architecture. It allows us to comprehensively understand the interplay among key design parameters - array size, aspect ratio, dataflow, and memory-bandwidth.
  \item Through a suite of case-studies, we demonstrate the impact of these parameters on performance, energy, and scalability across a diverse set of DNNs from MLPerf.
\end{itemize}

The rest of the paper is organized as follows.~\Sec{sec:background} introduces the background and motivation for this work.~\Sec{sec:impl} describes the simulation methodology.~\Sec{sec:case_studies} uses \tool to highlight design insights.
~\Sec{sec:RelatedWork} puts our work in the context of related work, and~\Sec{sec:conclusion} concludes.

\section{Motivation and Background}
\label{sec:background}


%

Given the large variety of applications for which deep learning based solutions
apply, there is naturally a broad spectrum of design points, from tiny low-power
embedded IoT devices through to large datacenter ASICs.
Regardless of where in the spectrum a design lies, it has to be practical. 
An embedded design should produce a result in seconds-not in hours; on the
other hand, a data center could not have a power plant of its own, so energy
consumption should be optimized even if the accelerator is designed for delivering high
performance.
Naturally these constraints pose several challenges when it comes to making design choices.  

Let's consider  Convolution Neural Networks (CNNs), which due to its relevance to a large gamut of problems, are one of the most widely used DNN layer types.
A convolution layer is a multi-dimensional kernel, and can therefore be mapped onto the hardware in more than one way, depending upon loop unrolling order.
The hardware itself, on the other hand, can schedule a given high level computational mapping in several possible ways. 
On top of this, there are system design choices like memory organization, interface design,
workloads scheduling etc which have profound implications on power and
performance.
Given the large number of factors at play, designing an efficient accelerator is a complex optimization problem of tuning multiple inter-related constraints.
The following sections describe these challenges in details.

\subsection{Mapping dataflows to the architecture}
\label{subsec:mapping}
A typical DNN accelerator comprises of multiple compute units which can perform
multiplication and addition (MAC).  
As mentioned earlier there any many possible ways of mapping the compute onto the array.  
Each such mapping is called a data-flow.  The data flows differ among themselves in terms
of latency, throughput and data reuse~\cite{eyeriss}. In the present
most of the DNN accelerators implement one data flow and optimizes around it.
~\cite{diannao2014asplos, pudiannao2015asplos, shidiannao2015isca, eyeriss}

However, not all data flows work efficiently for all types of layers in a given
CNN. In the past few years we have seen that CNN topology design trends
have changed a lot e.g., shallow networks with variable filter sizes (Alexnet), deep networks with fixed
filter sizes (VGG), 
depth wise separable
networks with variable filter size (Inception v4), 
and so on.  
An ideal case would be an accelerator which supports
variable data-flows. But such a design would incur some overhead and might not
be feasible.  
In such cases, it is important to keep the target application
topologies in mind at design time and optimize for the common case. 

\subsection{Hardware optimizations} \label{subsec:hardware_motivation} Assuming
that an optimal data flow is determined and agreed upon, there are still
multiple first and second order trade-off choices pertaining to final design
decision.  One of the first order choice is the layout of the compute units.
For maximum performance the compute units need to be kept fed and running at all
times.  Although this seems trivial, the problem compounds itself when the
number of compute elements become larger and the bandwidth of the communication
channels  cannot keep up. 
Moving to second order choices, the size of the on chip memory plays an major
role in power-performance trade-off.  If there is significant reuse, increasing
the chip memory can prove beneficial for performance.  However increasing memory
size also increases power consumption.  A small memory however will result in
spilling, increasing DRAM 
interface bandwidth requirement hurting both power and
performance goals.  Thus proper sizing of the memory is critical. 

Optimizing hardware by exploiting these trade-offs are at the heart of DNN accelerator research. For instance, prior works based on  dot-product
datapath~\cite{diannao2014asplos,dadiannao2014micro,shidiannao2015isca,pudiannao2015asplos} utilize reuse in only one dimension. While 2D arrays ~\cite{eyeriss,tpu-isca} utilize reuse in higher dimensions.    

\subsection{Simulators for DNN acceleration}
\label{subsec:sim_motivation}
There is no denying the fact that efficient design of DNN accelerator is a
key requirement to enable deployment of AI algorithms; be it in the embedded
setting or within a data center space.  Despite this importance, there has been
an surprising dearth of knowledge base and simulation-emulation infrastructure
in the public domain. 
A cycle-accurate simulation infrastructure for systolic arrays is lacking in the 
community today, which this work provides.
Related efforts for alternate accelerator architectures 
are discussed in Section~\ref{sec:RelatedWork}.

\section{SCALE-Sim: CNN Accelerator Simulator}
\label{sec:impl}

\tool is a simulator that provides a publicly available modeling
infrastructure for systolic array CNN accelerators.  \tool enables
designers to quickly iterate over and validate their upcoming designs with
respect to the various optimization goals for their respective implementation
points. In this section, we first describe the detailed modeling methodology,
including how we model the compute~(\Sec{subsec:compute_model}),
memory~(\Sec{subsec:memory_model}), and the accelerator's interface with the
whole system~(\Sec{subsec:interface_model}), respectively. 

\subsection{Modeling Compute} \label{subsec:compute_model}

\insertWideFigure{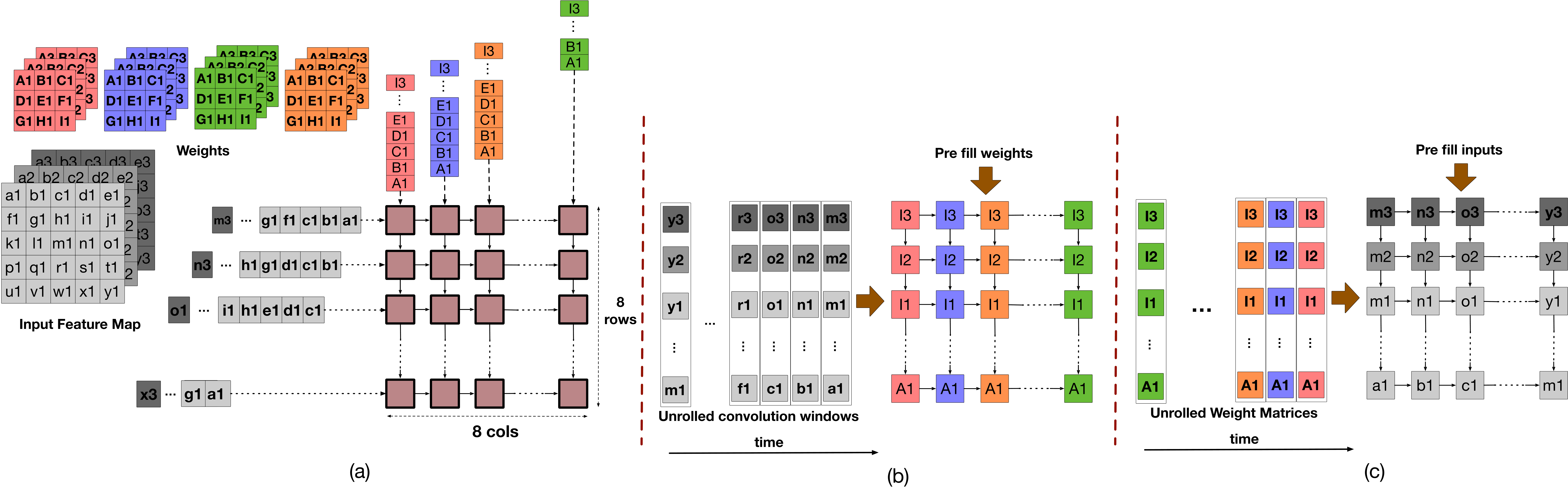}{ Schematic showing the mapping in various dataflows 
(a) Output stationary; (b) Weight stationary; (c) Input stationary}


\tool can simulate convolutions~\cite{imagenet, deepspeech} and matrix-matrix(MM) multiplications. Matrix-Vector(MV) and Vector-Vector(VV) multiplications are supported as special cases of MM with one or both dimensions as one. This lets us map fully-connected and recurrent layers in RNNs, which can be posed as MV problems.
We chose to support only these operations since, among the various compute operations performed for CNN inference, convolutions and fully connected layers comprise more than
90\%~\cite{eyeriss} of the total. 
Being the common case, it is therefore quite logical for accelerator
designers to optimize for the said operations. Optimizing
for any other step in general leads to diminished returns. 

\tool models the DNN accelerator's compute unit as a systolic array. Systolic
Arrays are effective, energy-efficient yet very simple designs to implement
matrix multiplication and similar operations on hardware.  As a result it is not
surprising that many DNN accelerator designs~\cite{tpu-isca, xdnn, eyeriss} are based
on systolic arrays.  The systolic array in DNN accelerators comprise of several
Multiply-and-Accumulate (MAC) units (also known as Processing Elements, or PEs),
bounded together in a two dimensional mesh.  The data is fed from the edges,
which then propagate to the elements within the same row and columns via
unidirectional links. Each MAC unit stores the incoming data in the current
cycle in an internal register and then forwards the same data to the outgoing
link in the next cycle.  This store and forward behavior results in significant
savings in SRAM read bandwidth and could very effectively exploit the reuse
opportunities provided by convolution operation, making this a popular choice
for accelerator design.

Among the recent accelerator designs, the shape of the systolic array has been almost invariably a square. 
Although depending upon the workload and mapping strategy this might not be the best shape for performance and energy efficiency. 
Acknowledging this possibility, \tool has the capability to take the length and breadth of the array as user input.   


\subsection{Modelling Dataflow}
\label{subsec:dataflow}

We define the term dataflow the same way as authors in Eyeriss do \cite{eyeriss}, and keep the same nomenclature.
A given mapping scheme determines the order in which inputs are
fetched, outputs are generated, and intermediate results are stored and reused. 
A particular dataflow determines the reuse within the array and the bandwidth requirements from the system.

Eyeriss describes
\textit{Output Stationary} (OS), 
\textit{WeightStationary} (WS), 
\textit{Input Stationary} (IS), 
\textit{Row Stationary} (RS), and 
\textit{No local reuse} (NLR).
Out of these, \tool models first three, which is provided by the user as input.
NLR is not explicitly modelled since it is a special case for any of the dataflows with small buffer memories. RS on the other hand is fundamentally tied to the PE design chosen in Eyeriss.
In the next few sections we will briefly describe these dataflows in the context of systolic arrays.

\noindent\underline{\textbf{Output Stationary (OS)}} \\
The stationary term in the name of the dataflow indicates the matrix which is "pinned" to a given PE. 
Output stationary therefore refers to the mapping where each pixel of the output feature map is assigned to a given PE. Given infinite resources, we would only need as many PEs as the number of output pixels. 
To achieve this, all the compute necessary for generating the given output
is done on the said PE and the required operands are streamed in every cycle. 
Reduction operation is done in place and no further communication is needed between the
MAC units as far as generating the given pixel is concerned. 
In a realistic case where the number of compute elements is limited, the resources are time multiplexed. 
Once one output pixel is generated by a given PE, the result is transferred to the memory
and the PE is assigned another pixel to compute.

\autoref{fig:dataflow}\textit{(a)} depicts the schematic of the dataflow in \tool. 
The data is fed from left and top edges of the array, where the left edges stream in input pixels while the top edge streams in pixels from the filter or weight matrices. 
In a given column PEs in each row are responsible for generating adjacent output pixels in a single channel. 
Each column however generates pixels corresponding to different output channels.

The dataflow model implemented in \tool assumes that 
the generated outputs can be transferred out of the array without incurring a stall in compute. 
In actual implementations this may not be true, hence the runtime might be higher than the calculated value.
The rationale behind this decision is that \tool depicts the opportunities and limitations arising from the dataflow itself, without any dependence on a specific implementation choice.

\noindent\underline{\textbf{Weight Stationary (WS)}} \\
Following in the convention of nomenclature mentioned above, \textit{Weight Stationary} dataflow refers to the mapping where each element of the weight matrix is uniquely mapped to a given MAC unit. 
Once a set of weights is mapped onto the array, they are not replaced until all the
computations involving the given set of weights is finished. 
Every cycle the input elements required to be multiplied with the currently mapped weights are
streamed and partial sums are stored within the array. 
Reduction takes place by communicating the partial sums across the MAC units present in the array and often takes multiple cycles. 
\autoref{fig:dataflow}\textit{(b)} shows the mapping in WS dataflow. 
The mapping takes place in two steps. 
First each column is assigned to a given filter. For a given column, the elements of the assigned filter matrix are fed in from the top edge, till all the PEs in the given column has one element each. 
After the filter elements are placed, the pixels of input feature map are then fed in from the left edge.
During this phase, the partial sums for a given output pixel is generated every cycle.
For a given output pixel, the corresponding partial sums are distributed over a column. 
These partial sums are then reduced over the given column in next $n$ cycles, where $n$ is the number of partial sums generated for a given pixel.
The weight pixels are kept in the array until all the computations which require these values as operands are not over. 
Once the computations corresponding to the mapped weight are done, the mapping is repeated with new set of weights. 

Due to the sequential nature of the mapping; first the weights are mapped and then the inputs are streamed. 
The number of SRAM banks needed to support this implementation is lower than output stationary implementation for a given array.
However, partial sums corresponding multiple output pixels are now required to be kept in the array, until they are reduced, which leads to increase in implementation cost.

\noindent\underline{\textbf{Input Stationary (IS)}} \\
\textit{Input Stationary} dataflow is a similar mapping as WS, where, as the name implies, pixels of the input feature map (IFMAP) are "pinned" with the PEs and elements of the weight matrices are streamed in. 

\autoref{fig:dataflow}\textit{(c)} depicts the schematic of the mapping. 
Similar to WS, this mapping also takes place in two stages. 
However, in this case, each column is assigned to a \textit{convolution window}. 
The \textit{Convolution window} is defined as the set of all the pixels in the IFMAP which are required to generate a single OFMAP pixel. 
As in the case of WS, for a given column the pixels corresponding to a given convolution window are streamed in from the top edge. 
Once the input pixels are fed in, the elements of the weight matrices are streamed in from the left edge. 
Again similar to WS, reduction is performed over a given column, and the convolutions windows are kept around until all the computations requiring these elements are done, before remapping the array with elements belonging to new convolution windows. 

This dataflow also enjoys the benefits of lower SRAM bank requirements, as compared to OS. However the cost and runtime compared to WS varies by workload.

\subsection{Modeling Memory}
\label{subsec:memory_model}


CNNs are memory-intensive. 
Therefore, the memory hierarchy design
is critical to the overall performance and energy consumption of a CNN
accelerator. However, determining the optimal memory system design is
non-trivial as the memory system must be co-designed with the compute array. \tool models a parameterizable memory hierarchy for CNN accelerators and
allows for co-optimizations between the computer units and the memory hierarchy.

The key to the on-chip memory hierarchy design for CNN accelerators is to exploit the ample data reuse provided by the convolution operations. A typical convolution
can be viewed as a small filter kernel being slid over a given input matrix,
with each overlap generating one output pixel.  When the convolution operation is
formulated as successive dot-product operations, three reuse patterns
are immediately evident. 

\begin{itemize}
  \item Each convolution window uses the same filter matrix, to generate pixels corresponding to a given output channel
  \item The adjacent convolution windows share portions of the input matrix if the stride is smaller than window dimension
  \item To generate a output pixel in different output channels, different filter matrices use the same convolution window.
\end{itemize}
In short there are spatial, temporal and
spatio-temporal reuse patterns in CNN inference. 

The memory system design would ideally keep the working set
operands near the compute elements in any DNN accelerator. Naturally almost all
accelerator designs provision for some scratchpad memory.  However, determining
the size of the scratchpad memory is non-trivial.
\autoref{subsec:hardware_motivation} provides an insight on how memory size
affects power and performance.
Further complexity arises as the reuse behavior heavily
depends upon the dataflow and the DNN layer hyper-parameters.  Hence,
determining optimal size of the memory is empirical and requires simulating the
accelerator behavior when target workloads are used. 

In \tool we model the memory in three logical partitions that store IFMAP, filter matrices and the generated OFMAP, respectively. 
The size of each partition is user-specified.
The utility for input and weight partitions are obvious, but at the first glance the output partition seems to be redundant.
The output partition, however, serves two purposes. First, it stores
the outputs till there are enough elements to allow for bursty transfers. Second, in case of WS an IS implementations it stores the partial sums.

All the memories in \tool are modeled as double buffers to hide the SRAM access latency which a standard practice. 
For the uninitiated in the double buffered implementation there are two sets of memories, 
a working set and an idle set. 
At any given time the working set is used to feed the array and
the idle set is populated by fetching data from the off-chip
memory. 
For outputs, the working set is populated by the compute while the idle
set is used by the DMA or equivalent memory controller for transferring its
contents back to the off-chip memory.

\subsection{Modeling System Interface} \label{subsec:interface_model}
\insertFigure{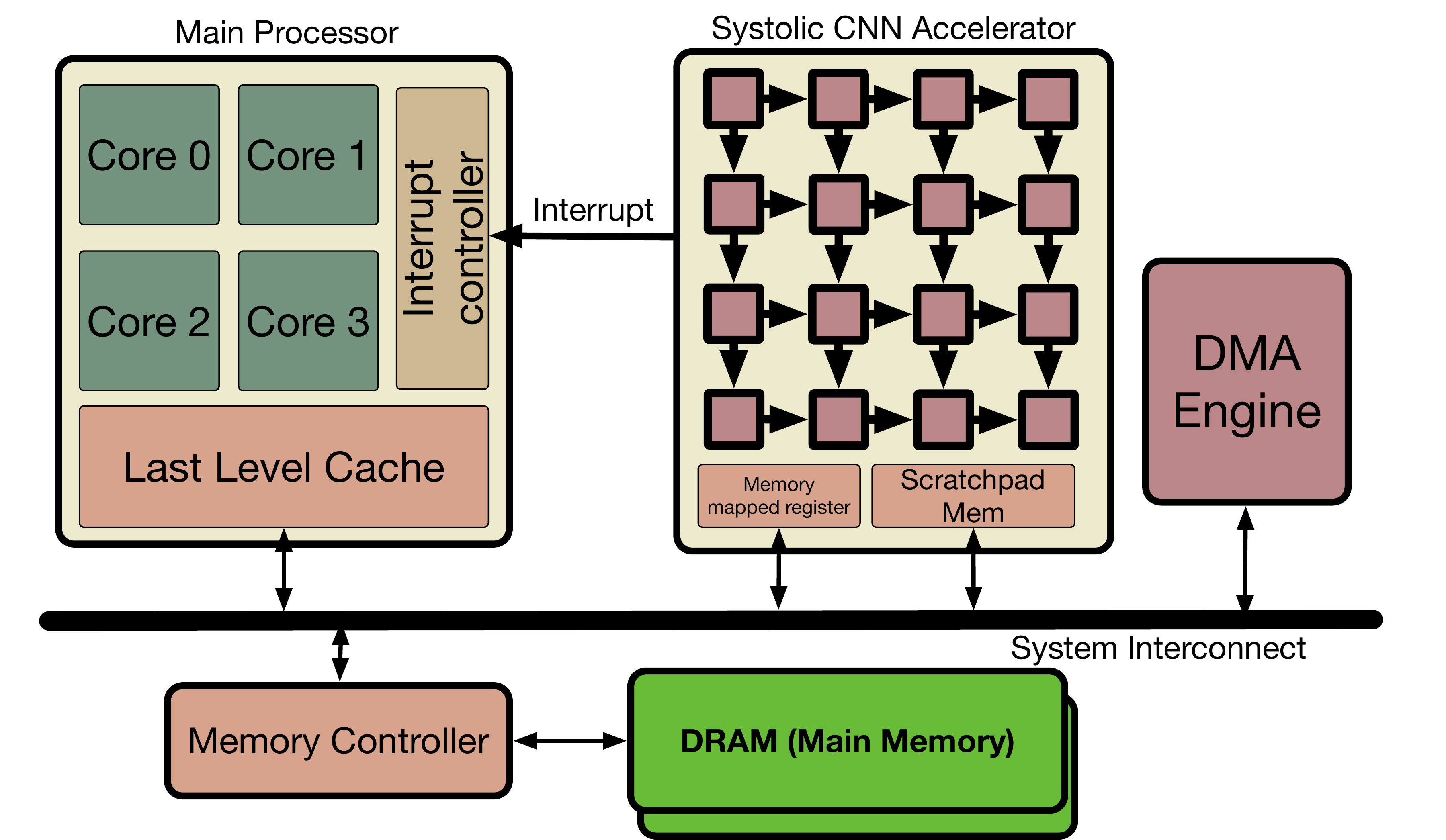}{Schematic showing the integration model of accelerator in a systems context}

An accelerator by definition is a co-processing element augmented with a main
processing system to improve overall performance.  However, contemporary pieces
of work in this space tend to overlook the system integration aspect assuming
that the given accelerator design will integrate as is.  However, aggressive design points leading to optimal accelerator performance might result
it suboptimal system performance. For example, an accelerator design might have
multiple processing elements to exploit parallelism, but in reality system
memory is unable to supply enough operands to keep all the units busy~\cite{gem5aladdin}. It is therefore important
to understand the implication of integrating an DNN accelerator into the overall system.

We consider the typical model of accelerator integration in \tool. That is to attach the DNN accelerator to the system
interconnect on a slave interface as \autoref{fig:system} shows. 
The master is a processor which interacts with the accelerator by writing task descriptors to
memory mapped registers.  
When a task is offloaded to the accelerator, the master context switches to work on other jobs, while the accelerator wakes up and starts computing, independently generating its memory
requests and side channel signals. 
When finished, 
the accelerator then copies the result to the memory and notifies the master. 

With this interaction model in mind, it could be seen that the interface with the system bus serves as a reasonable proxy for the overall behavior when it is integrated with the system.  \tool allows for modeling the main memory behavior by generating accurate read and write bandwidths of the interface, which can then be fed into a DRAM simulator eg. DRAM-Sim2\cite{dramsim2}.
In general, as described in \autoref{subsec:memory_model} the memories inside a DNN
accelerator are scratchpad memories and not caches.  Therefore, coherence is not
managed in hardware. Thus, we do not model it in \tool. 

 
\subsection{Implementation}
\label{subsec:simulation}

\begin{table}[t] 
\centering 
\setlength{\abovecaptionskip}{3pt}
\setlength{\belowcaptionskip}{0pt} 
\scriptsize \begin{tabular}{|p{2cm}|p{6cm}|}

\hline \bf Parameter & \bf Description \\  
\hline ArrayHeight & Number of rows of the MAC systolic array \\ 
\hline ArrayWidth & Number of columns of the MAC
systolic array \\ 
\hline IfmapSRAMSz & Size of the working set SRAM for IFMAP in KBytes \\ 
\hline FilterSRAMSz & Size of the working set SRAM for filters in KBytes \\ 
\hline OfmapSRAMSz  & Size of the working set SRAM for OFMAP in KBytes\\
\hline IfmapOffset & Offset to the generated addresses for IFMAP px \\ 
\hline FilterOffset & Offset to the generated addresses for filter px \\
\hline OfmapOffset & Offset to the generated addresses for OFMAP px \\
\hline DataFlow & Dataflow for this run. Legal values are 'os','ws', and 'is' \\
\hline Topology & Path to the topology file \\ 
\hline 

\end{tabular}
\caption{\tool config desciption} 
\label{table:config} 
\end{table}

\tool internally takes an inside out implementation approach. 
Specifically, the simulator assumes that the compute units are always used to the maximum possible utilization - as dictated by dataflow mapping, and never stall waiting for data. 
As mentioned before, the rationale behind this choice is that \tool highlights the 
opportunities and limitations by virtue of the modelling parameters (eg. dataflow, memory sizing) and not any specific implementation choice.
With this implementation model, the simulation in \tool takes place in following steps. 
\begin{itemize}
\item { \tool generates cycle accurate read addresses for elements required to be fed on the top and
left edges of the array \textit{such that the PE array never stalls}. These address are effectively the SRAM read traffic for filter and input matrices, as dictated by the dataflow. Given the reduction takes predictable cycles after data has been fed in, \tool generates output trace for the output matrix, which essentially constitutes the SRAM write traffic.}
\item { \tool parses the generated traffic traces, to determine total runtime for
compute and data transfer to and from SRAM. The data transfer time is essentially the cycle count of last output trace entry. Parsing the SRAM traffic also provides utilization information of the array.}
\item { Given the SRAM traffic and the SRAM configurations, \tool then generates DRAM traffic trace, for both input and output data transfer.}
\item{ Finally the DRAM traces area parsed to estimate the memory bandwidth requirement as well as memory power consumption.}
\end{itemize}

\insertFigure{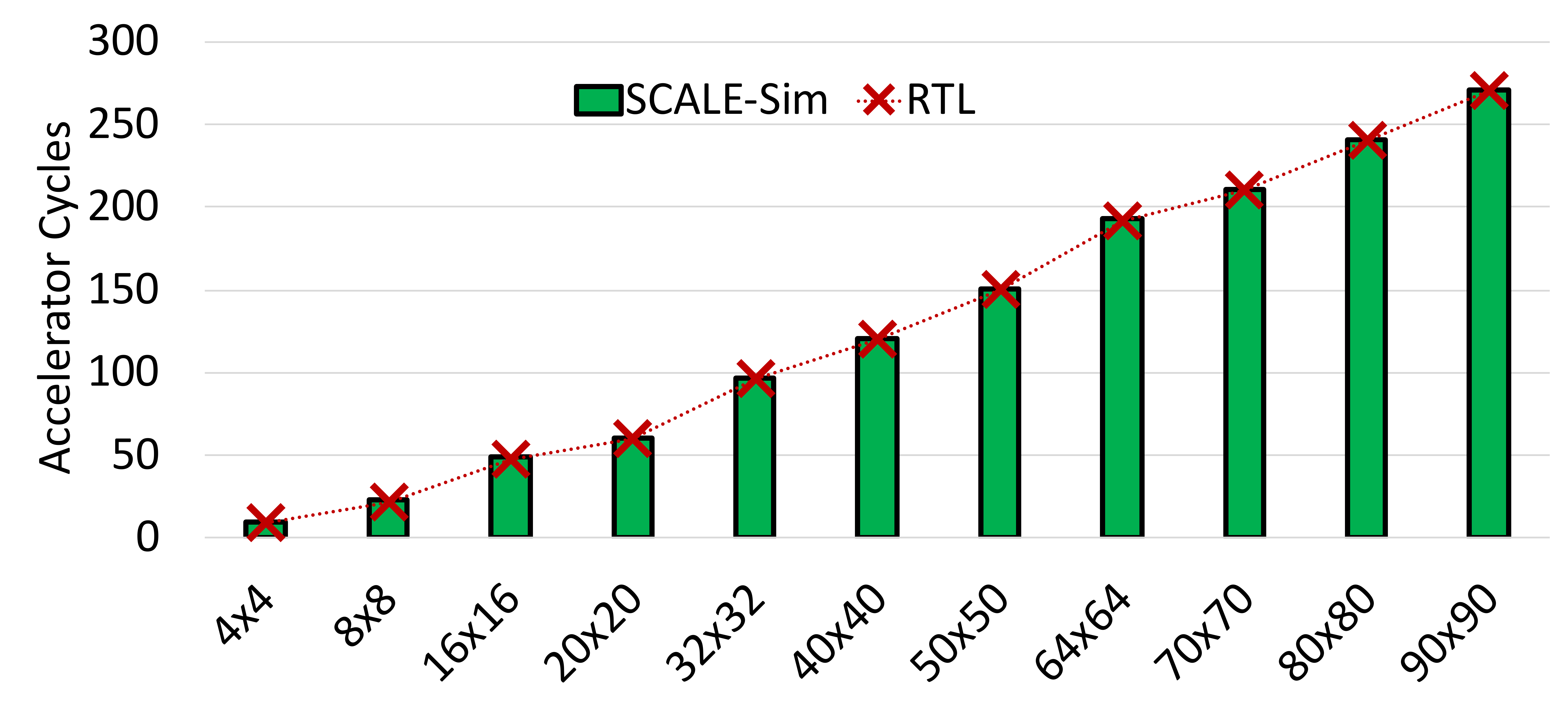}{Cycles reported by \tool runs and RTL simulation for Mat-Mat multiplication workloads}

\underline{\textit{Validation}}
We validate \tool against an in-house RTL model for a systolic array implementing OS dataflow. The workload we chose is a Mat-Mat multiplication of matrices with same size as the array. \autoref{fig:validation} depicts how the run times tally for the two platforms.

\subsection{User Interface}
\label{subsec:ui}
\vspace{-2mm}

\begin{table}[t] 
\centering 
\setlength{\abovecaptionskip}{3pt}
\setlength{\belowcaptionskip}{0pt} 
\scriptsize \begin{tabular}{|p{2cm}|p{6cm}|}

\hline \bf Parameter & \bf Description \\  
\hline Layer Name & User defined tag \\ 
\hline IFMAP Height & Dimension of IFMAP matrix \\
\hline IFMAP Width & Dimension of IFMAP matrix \\
\hline Filter Height & Dimension of one Filter matrix \\
\hline Filter Width & Dimension of one Filter matrix \\
\hline Channels & Number of Input channels \\
\hline Num Filter & Number of Filter matrices. This is also the number of OFMAP channels \\
\hline Strides & Strides in convolution \\
\hline 

\end{tabular}
\caption{\tool Topology file desciption} 
\label{table:topo} 
\end{table}

The tool takes two files as inputs from the user, a config file and other is a topology file. 
This file contains the user specification for architectural parameters, like the array size, the memory size etc and the path to the topology file. 
\autoref{table:config} depicts the complete list of parameters.

The topology file contains the information of the hyper-parameters for the various layers in a given DNN. This is a csv file, with each row listing all the required hyper-parameters for a given layer, \autoref{table:topo} gives the complete list of all the entries in a given row. 
\tool parses the topology file one line at a time and simulates the execution of the layer. This is a natural model for traditional DNNs. However, in modern DNNs there are cells with multiple conv layers in parallel ~\cite{resnet}. \tool serializes the execution for such layers in the same order in which the layers are listed in the topology csv file.

\tool generates, two types of outputs. First is the cycle accurate traces for SRAM and DRAM reads and writes. The traces are also csv files, which list the cycle and the addresses of data transferred in the given cycle. The other type of output file are the metrics files, which summarizes the parsed information from the traces, these include cycle counts, utilization, bandwidth requirements, total data transfers etc.

\section{Design Insights using \tool} \label{sec:case_studies}

\insertWideFigure{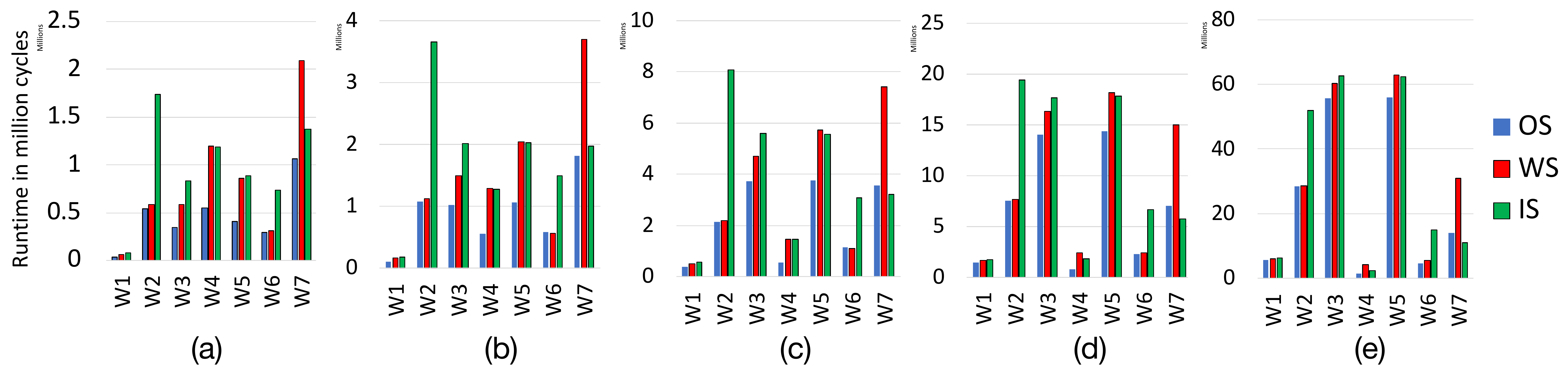}{Chart showing the runtime in cycles to compute all the layers of our workloads while using different dataflows in square arrays with the dimensions (a)128x128, (b)64x64, (c)32x32, (d)16x16, and (e)8x8 \vspace{-2mm}}
\insertWideFigure{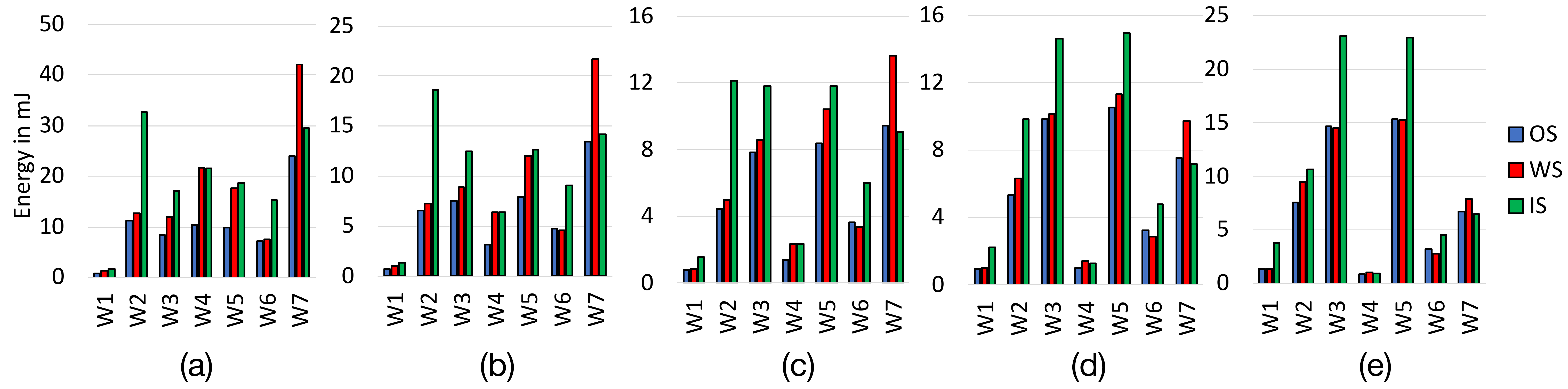}{Chart showing the energy consumed in mJ, in compute and memory transfers for all the layers in our workloads using different dataflows in square arrays with dimensions (a)128x128, (b)64x64, (c)32x32, (d)16x16, and (e)8x8 \vspace{-2mm}}

Systolic array is a well known architecture for implementing matrix multiplication
and several other linear algebra kernels in hardware.
The communication of operands in the array is very efficient as the data is only
passed between neighbors and does not require any global communication.
Address generation and matching is also not required.
Being a natural fit, there have been several systolic-array based CNN accelerators, with some used in commercial applications, like the Google TPU~\cite{tpu-isca} and Xilinx xDNN \cite{xdnn}.

However, the design process for 2D arrays is not straight-forward due to the large
number of free parameters and the variety of DNN workloads in common use.
We use \tool to observe the effects of various design decisions on the performance and energy of a systolic-array based accelerator.

In the next few sections we will describe and present results from various experiments performed using \tool targeted towards efficient accelerator design.  

\subsection{Methodology} 
\label{subsec:method}

\begin{table}[t] 
\centering 
\setlength{\abovecaptionskip}{3pt}
\setlength{\belowcaptionskip}{0pt} 
\scriptsize \begin{tabular}{|p{2cm}|p{6cm}|}

\hline \bf Tag & \bf Description \\  
\hline W1 & AlphaGoZero \cite{alphagozero} \\ 
\hline W2 & DeepSpeech2 \cite{deepspeech2} \\ 
\hline W3 & FasterRCNN \cite{fasterrcnn}\\ 
\hline W4 & Neural Collaborative Filtering \cite{ncf}\\ 
\hline W5 & Resnet50 \cite{resnet}\\
\hline W6 & Sentimental CNN \cite{sentiment}\\ 
\hline W7 & Transformer \cite{transformer} \\
\hline 

\end{tabular}
\caption{Workloads used in this work} 
\label{table:workloads} 
\end{table}

As mentioned in the previous sections, \tool gives the user to choose the micro-architectural parameters.
In the following set of experiments we study the affect of each of these parameters by sweeping over a range of values while keeping others constant. 
Unless stated otherwise, we keep the number of compute units same as that of TPUv3 (128x128 MACs). 
However unlike TPUv3 we assume a data size of 1 byte, which is standard for DNN inference. 
The default size of scratchpad memory for operands in our experiments in 1024 KB, with 512 KBs allocated to filter and IFMAP buffers each.
To ensure that our workloads represent a wide class of applications in the ML space, we chose MLPERF \cite{mlperf}. 
MLPERF is an ongoing effort in the machine-learning and systems community to provide a standard benchmark for both software and hardware frameworks. \autoref{table:workloads} shows the workloads we use. The user might notice that we have replaced Masked-RCNN with FasterRCNN, since both these workloads have similar hyper-parameters but FasterRCNN finishes faster in our simulations.

\subsection{Effect of dataflow}
\label{subsec:effect_df}

\insertWideFigure{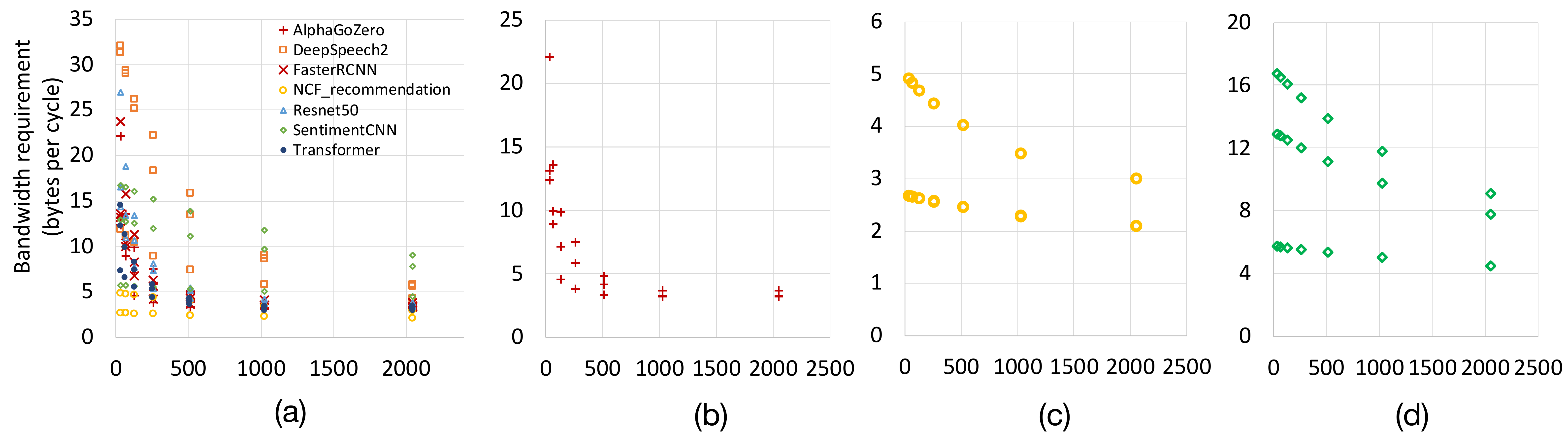}{Figures depicting the required DRAM bandwidth vs scratchpad memory size for stall free operation for (a) all workloads, (b) AlphaGoZero(W1), (c) Neural Collaborative Filtering(W4), (d) Sentimental analysis CNN (W6) \vspace{-2mm} }

As we have discussed in \autoref{subsec:dataflow} \textit{dataflow} refers to the mapping strategy of compute on the array.
It is not surprising that dataflow will have direct implications on the performance and efficiency of the system.
Most of the recent DNN accelerators chose a particular dataflow and stick with it. 
The microarchitecture is usually optimized around this dataflow thus rendering the mapping scheme immutable. 
The authors of FlexFlow~\cite{flexflow} argue that this practice leads to inefficiencies in terms of energy and performance and hence propose a custom accelerator to combat the same. 

In this section we use \tool to see if the argument holds in case of systolic arrays. In particular we try to answer the following questions, 
\begin{enumerate}
\item{ \textit{Does the size of the array dictate the choice of dataflow?}}
\item{ \textit{How much does the hyper-parameters of the workload dictate choice of dataflow?}} 
\item{ \textit{Are we missing out a lot by employing fixed dataflows? Or is there a dataflow which works in all cases?}}
\end{enumerate}

\autoref{fig:runtime_vs_dataflow} and \autoref{fig:energy_vs_dataflow} depict the trends in runtime and energy when arrays of different size run the workloads with various dataflows. In a glance it seems, \textit{Output Stationary}(OS) outperforms the other two dataflow in every aspect. However, it might be worth noting that implementing a stall free OS hardware might not be trivial. Furthermore in the energy calculations, the cost of logic within the accelerator is assumed to be the same for the three dataflows, which might not always hold true. Moreover, for square arrays, WS and IS use half the amount of SRAM banks as compared to OS. SRAM banks are expensive resources in terms of area footprint.
 
Among IS and WS we see interesting trends, which help us address the questions we raised before. 
Looking at the runtime trends for W4, we notice that for larger array sizes, 128$\times$128 and 64$\times$64 the IS and WS performance is comparable. However as the array sizes decrease, IS turns out to be more performant than WS. This helps us answer the first question. If W4 or similar networks are dominant workloads for a given use case, then the choice of dataflow is tied with the size of the array to extract maximum performance.   

Transferring our attention to W2 and W7, we notice that WS and IS are clear winners in the respectively in these workloads. This trend is invariant of the size of the array and therefore is clearly dependent on network hyper-parameters. In general for WS and IS dataflows, the less times the 'stationary' matrix is needed to be mapped into the array, the better. This is because mapping of stationary matrices take cycles which cannot be utilized for compute. If in a layer the number of output pixels are larger the the number of weights then WS will outperform IS and vice versa. 

Although we observe clear indications that hyper-parameters and array size affects the choice of dataflow, the trends in \autoref{fig:runtime_vs_dataflow} and \autoref{fig:energy_vs_dataflow} do not show dramatic effects, unless designing a highly optimized design tied to any workload. Therefore to answer question 3, although there might not be a dataflow choice to rule them all, fixating to a given dataflow might not lead to significant losses in terms runtime or energy. Therefore while designing systems with flexible dataflows, the cost of implementation should be carefully evaluated.

\vspace{-2mm}
\subsection{Effect of Memory Sizing}
\label{subsec:effect_mem}
\vspace{-2mm}


Providing sufficient on-chip memory has a major implication of the performance of the DNN accelerators. Due to the massive reuse opportunity provided by CNNs, providing large enough memory can significantly reduce off-chip accesses and hence improve the energy consumption and overhead on the system. However memory is an expensive resource both in term of area and power consumption. Therefore, sizing the memory appropriately is an important aspect for accelerator design. 

\autoref{fig:memory_sweep} shows the off-chip bandwidth requirement for various worklaods when the on-chip memory size is increased from 32KB to 2048KB for each Filter and IFMAP buffers. \autoref{fig:memory_sweep}(a) shows that the return diminish after hitting 1MB buffer size for the common case. However, for specific workloads the knee of the curve varies significantly. \autoref{fig:memory_sweep}(b) shows that W1 hits the knee at 256KB, while \autoref{fig:memory_sweep}(c) shows for W4 the knee lies for even smaller sizes. On the other hand W6 shows improvements even after 1024KB as depicted in \autoref{fig:memory_sweep}(d)

\vspace{-2mm}
\subsection{Effect of Shape of the array}
\label{subsec:effect_aspect}
\vspace{-2mm}

\insertFigure{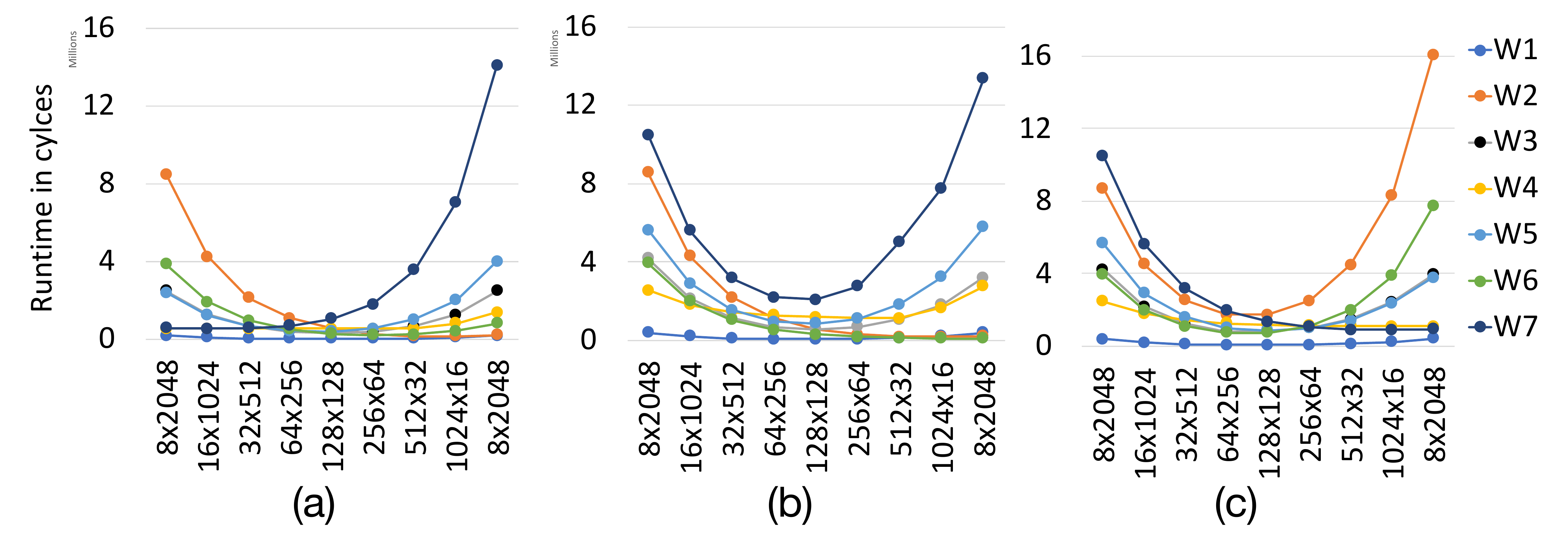}{Chart showing the variation of runtime for our workloads run different shapes of systolic array with fixed number of PEs (=16384) for three dataflows (a) Output Stationary, (b) Weight Stationary, (c) Input Stationary }

Given a fixed number of compute units, a set of hyper-parameters and a given dataflow, it is quite interesting to explore the effect of the shape of the array on performance. We use \tool to compute runtime for our workloads by changing the shape of systolic array from, 8x2048 to 2048x8. \autoref{fig:aspect_ratio_runtime} shows our findings. 
It is quite interesting to observe that the combination of dataflow and shape has dramatic trends. For instance OS and WS favor short-wide configurations for W4, while IS favors square aspect-ratios. On the other hand, OS and IS favor completely different configurations for W7. Interestingly square aspect ratios perform well for the common case.    

\vspace{-2mm}
\subsection{Scaling out vs Scaling up}
\label{subsec:scale_up_vs_out}
\vspace{-2mm}

It is a no-brainer that performance can be increased by throwing more compute elements on a highly-parallelizable workload, such as CNN inference. However, scaling can achieved in two major ways, by adding making the array bigger \textit{(scaling-up)} or by having more arrays and dividing compute among them \textit{(scaling-out)}. 
The TPU is an embodiment of the former approach, while NVIDIA's tensor cores of the latter.

\insertFigure{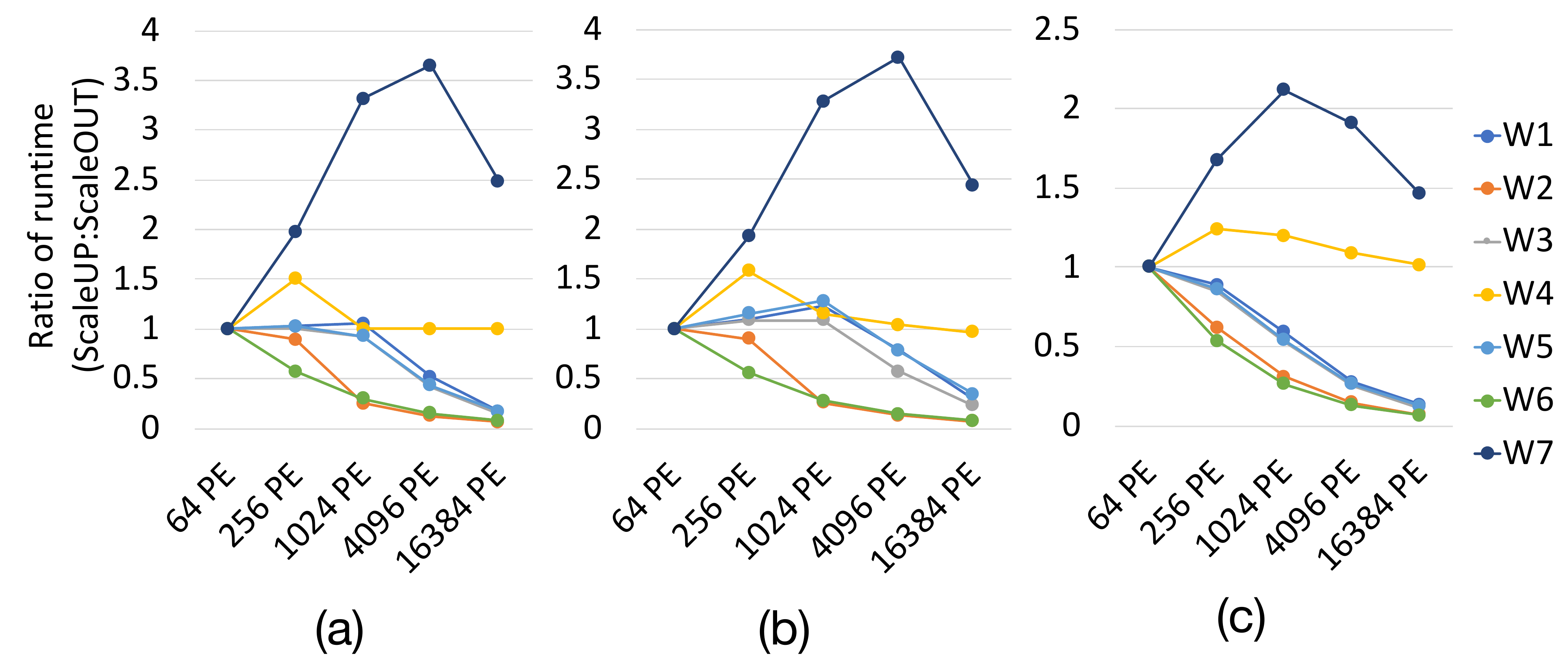}{Figure showing the ratio of runtime between runs on a scaled-up array vs a scaled-out implementation where the total number of compute elements are equal for different dataflows (a) Output stationary, (b) Weight stationary, (c) Input stationary.}
\insertWideFigure{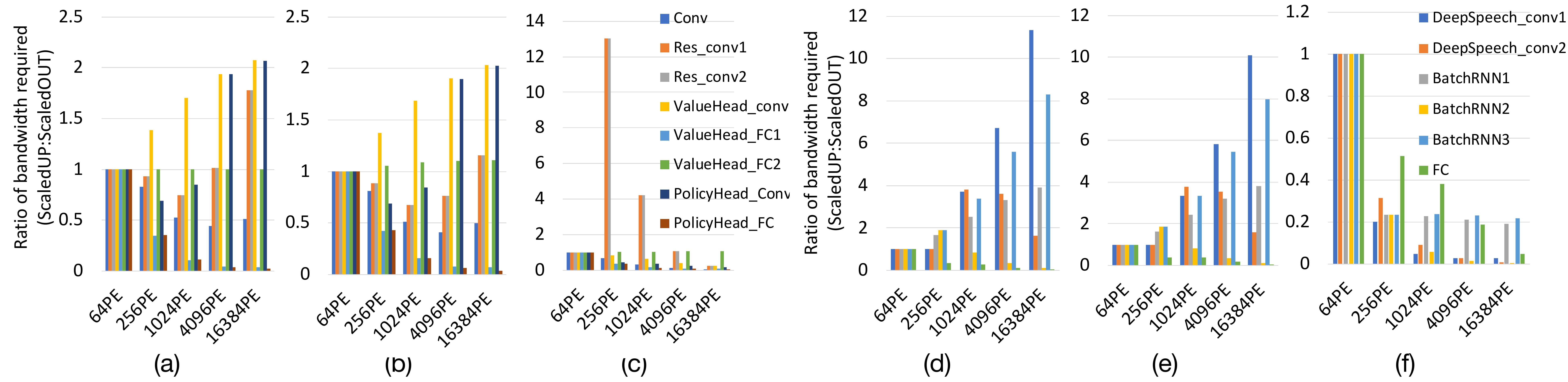}{Chart depicting the ratio of DRAM bandwidth requirement for weight matrices between runs on a scaled-up array vs a scaled-out implementation for the three dataflows for AlphaGoZero(W1)(a-c) and DeepSpeech2(W2)(d-f)}

We performed an experiment to explore the tradeoffs between the two approaches. 
We start from an 8x8 array (64PEs) and increase the number of compute elements to 16384, multiplying by 4 in each step. For scaling-up each step corresponds to doubling the length and breadth of the array, while for scaling-out each step is just quadrupling the number of 8x8 arrays. 
For the purposes of this case study, we divide the workload for scale-out along the output channels, i.e., the different filters are assigned to different nodes thus different nodes generating different output channels. Alternate partitioning strategies exist, and in fact the best strategy may differ from layer to layer depending on the number of filters vs channels. We do not add any arbitration or bandwidth constraints on the interconnect connecting the  arrays in the scale-out mode; instead \tool's outputs for SRAM read bandwidth requirement determines the bandwidth requirement for this interconnect.

\autoref{fig:scale_up_to_out_runtime} shows the ratio of scaled-up to scaled-out runtime for the three dataflows. For the common case scaled-up implementation turns out to be the best in terms of performance, ie. runtime(scale-up) $<$ runtime(scale-out). However, W1 favors scale-out irrespective of dataflow, indicating that scaling decision to be tied to workloads. This is a very important observation for designing high performance machine targeted for specific workloads. 

We also study the system level affect of these two scaling alternatives. Specifically we study the affect on DRAM bandwidth requirements for filter weights.
\autoref{fig:scale_up_to_out_bw} shows the ratio of DRAM requirements for scale-up to scale-out for each layer in AlphaGoZero (W1) and DeepSpeech2 (W2). For OS and WS in W1 (\autoref{fig:scale_up_to_out_bw}(a-b) respectively) we see most of the layers favor scaled-up implementation (bandwidth(scaled-up) $<$ bandwidth(scaled-out)). 
However, as the number of PEs increase the trend shifts towards scaled-out implementation.
However, for IS \autoref{fig:scale_up_to_out_bw}(c), the trend is reversed; for smaller PE counts (256PE, 1024PE) certain layers prefer scaled-out, while as PEs increase, scaled-out implementation wins. 
For DeepSpeech2 (W2) we see similar trends (see \autoref{fig:scale_up_to_out_bw}(d-f)). IS in W2 however, strongly favors scale-up as seen in \autoref{fig:scale_up_to_out_bw}(f).

\vspace{-2mm}
\section{Related Work}
\label{sec:RelatedWork}
\vspace{-2mm}

\textbf{Algorithmic Optimizations}.
CNNs are amenable to a range of optimizations on top of more fundamental
architecture decisions.  Prior work has explored exploiting optimized
datatypes~\cite{gupta-2015,minerva,stripes2016micro},
operand sparsity~\cite{eie,minerva,scnn}, and hardware faults~\cite{minerva}.
These optimizations are largely orthogonal to this work.


%
%
%
%
%
%
%

\textbf{Simulators}.
\tool is the first public and open source CNN accelerator that we are aware of.
However, there are a number of related tools and simulation methodologies that
we should be mentioned here.
Aladdin~\cite{aladdin} is a tool for simulating power, performance and silicon
area of arbitrary accelerators.
The methodology follows an HLS-inspired approach that starts with a C-code description of the algorithm.
This description is parsed into an LLVM graph, which is then scheduled into a
hardware pipeline,
guided by some simple constraints to describe the degree of parallelism and the
memory bandwidth.
This approach is ideal for rapid exploration of the hardware cost of a range of
algorithms.  However, it is somewhat limited in the sophistication of hardware
structures that can be generated.
Minerva~\cite{minerva} builds on top of Aladdin and uses a customized neural network training flow to
explore hardware-algorithm co-design opportunities. However, Minerva does not provide architectural insights such as resource underutilization or main memory bottleneck as \tool does.

Some papers like SCNN~\cite{scnn} and energy-aware pruning~\cite{vivienne-pruning} have mentioned and introduced some power measurement and
simulation infrastructure but either the scope of generalization is highly
limited or they are unavailable to the public. 
Alternately, highly custom designs like MAERI~\cite{maeri} have been released as open-source 
RTL models, 
but require a buy-in to a highly configurable design
that adds significant area and power overheads over 
simple systolic arrays. RTL simulation is also much slower than \tool's python-based 
cycle-accurate  model.
Tetris \cite{tetris} provides a tool to partition and schedule NN layers over Eyeriss \cite{eyeriss}, but unlike like \tool it is not a simulator.

Finally, we note that usually in an ML-enabled application, CNN is just one stage of the end-to-end
processing pipeline~\cite{euphrates, mobilecpu}. Therefore, it is important to understand the accelerator in the context of the entire Systems-on-a-chip (SoC). Gem5-Aladdin~\cite{gem5aladdin} embeds the Aladdin accelerator simulator inside the Gem5
system simulator environment to allow for system trade-offs to be explored. GemDroid~\cite{gemdroid} couples Gem5 with the Android Emulator and integrates various hardware IP models to enable SoC-level simulation. However, both pieces of work lack accurate account for DNN-specific accelerators, and therefore are not readily available for studying system-level
behavior of ML-enabled applications.

%

Due to the modular interface design, users could choose to integrate \tool with
Gem5-Aladdin or GemDroid for full-system simulation. This is particular helpful for
researchers who do not wish to perform in-depth investigation of the CNN
accelerator microarchitecture, but wish to integrate a decent CNN IP to perform
meaningful system-level characterizations.



\vspace{-2mm}
\section{Conclusion}
\label{sec:conclusion}
\vspace{-2mm}
 
In light of the fact that more and more architects are now designing
accelerators for deep neural networks, it is striking that there is a dearth of
publicly available knowledge base or simulation infrastructure to study design
insights.  In this work we make an attempt to bridge this gap in two ways. First
we implement a simulation tool for conducting our study and open source it for the general public. Second we perform detailed experiments to understand the design space and trade-off studies in designing a systolic array based CNN accelerator.  
We hope that our findings will help speedup development of new
accelerator designs and our tool help people conduct their design space explorations faster.

\bibliographystyle{IEEEtran.bst}
\bibliography{scale}

\end{document}